\documentclass[12pt,a4paper]{article}
\usepackage{amsmath}
\usepackage{amssymb}
\usepackage{amsthm}
\usepackage{float}
\usepackage{amsfonts}
\usepackage{graphicx}
\usepackage{verbatim}
\usepackage[left=2cm,right=2cm,top=3cm,bottom=2.5cm]{geometry}
\usepackage[numbers]{natbib}
\usepackage[utf8]{inputenc}
\usepackage[usenames,dvipsnames,svgnames]{xcolor}
\usepackage[colorlinks=true,
      linkcolor=red,
      urlcolor=gray,
      citecolor=blue]{hyperref}

\def\myalign#1{%
  \def\trule{\noalign{\smallskip\hrule\medskip}}
  \def\nebc{\nearrow\bigcup}
  \def\sebc{\searrow\bigcup}
  \def\pminf{{}_{-\infty}|^{+\infty}}
  \let\Inf\infty
  \def\amp{&} 
  \vbox{\mathsurround0pt\openup1\jot
    \halign{%
      &$\displaystyle##\hfil\tabskip0pt$&\amp##\tabskip1em\crcr
      \noalign{\hrule height1pt\smallskip}#1\noalign{\smallskip\hrule height1pt}\crcr}}}
      
\begin{document}
\begin{center}
\textbf{Large-scale structure power spectrum from scalar-tensor gravity}
\end{center}
\hfill\\
Joseph Ntahompagaze$^{1}$, Amare Abebe$^{2}$ and Manasse R. Mbonye$^{1,3,4,5}$\\
\hfill\\
$^{1}$ Department of Physics, University of Rwanda, Kigali, Rwanda\\
$^{2}$ Center for Space Research, North-West University, Mahikeng, South Africa\\
$^{3}$ ICTP-East Africa Institute for Fundamental Research,  Kigali, Rwanda\\
$^{4}$ Rwanda Academy of Science, Kigali, Rwanda\\
$^{5}$ Rochester Institute of Technology, NY, USA\\
\hfill\\
Correspondence: ntahompagazej@gmail.com\;\;\;\;\;\;\;\;\;\;\;\;\;\;\;\;\;\;\;\;\;\;\;\;\;\;\;\;\;\;\;\;\;\;\;\;\;\;\;\;\;\;\;\;\;\;\;\;\;\;\;\;\;\;\;\;\;\;\;\;\;\;\;\;\;\;\;\;\;\;\;\;\;\;\;
\begin{center}
\textbf{Abstract}
\end{center}
This work deals with the computation of the power spectrum of large-scale structure using the dynamical system approach for a 
multi-fluid universe in scalar-tensor theory of gravity.
We use the $1+3$ covariant approach to obtain evolution equations and study 
 the behavior of the matter power spectrum of perturbation equations.
The study is based on the equivalence between $f(R)$ theory of gravity and
scalar-tensor theory of gravity. We find that, for power-law $(R^{n})$ models, with $1<n<1.3$, 
we have the power spectrum evolving above general relativistic scale-invariant line. For $n\geq 1.3$, the power spectrum starts with constant amplitude 
then it experiences oscillations and eventually saturates at finite amplitude. 
Such  behavior is consistent with other observations in the literature. 
The result supports the ongoing investigations of the equivalence between $f(R)$ and scalar-tensor theory at linear order.\\
\hfill\\
\textit{keywords:} $f(R)$ gravity --- scalar-tensor --- scalar field ---cosmology ---covariant perturbation

\section{Introduction}\label{INTRODUCTION}
In studies of large scale structure based on General Relativity (GR) the matter power spectrum has been shown to 
be scale invariant \cite{dunsby1992covariant,arbuzov2009general}.
The modification of GR leads to different theories of gravity. In studying matter power spectrum in such theories one seeks to know whether, or not, among other observational constraints,  the behavior of that power spectrum would be scale invariant.
Several studies have been done in the context of analyzing the behavior of the matter power spectrum in $f(R)$ models.  In \cite{geng2015matter}, the authors considered linear perturbations in two different types $f(R)$ models: power law and exponential law. They pointed out that amplitude of the matter power spectrum resulting from power law $f(R)$ model is bigger than that of
exponential law $f(R)$ models. The dependence of energy density perturbations are scale-dependence for both models.
In \cite{li2012non}, the power spectrum is computed using N-body simulation method for $f(R)$ models proposed in
\cite{li2011chameleon} and the scale-dependence of the power spectrum has been shown. Several $f(R)$ models have been treated in \cite{SanteCarloni1}, where power spectrum was obtained through 
the use of the $1+3$ covariant approach to establish the matter energy density dependence on wave-mode $k$. In \cite{abebe2015breaking}, power spectrum computations were used to distinguish between two models, GR and a re-constructed $f(R)$ model.
In the present paper, based on the work done in \cite{ntahompagaze2017f}, we relate scalar field and $f(R)$ theory to
study power-spectrum in scalar-tensor theory and use the results to develop the perturbation equations.

\noindent In studying the matter power spectrum of the energy density perturbations, one can introduce dynamical system techniques  \cite{26} in the analysis of cosmic evolution equations. 
The working scheme can be summarised as follows.
One defines dimensionless variables rendering the field equations  dimensionless \cite{26, bahamonde2018dynamical, abebe2013large}.
The result can be used in perturbation equations to construct the matter power spectrum \cite{abebe2013large}. 

\noindent  We use the above scheme to compute matter power spectrum in scalar-tensor theories of gravity. 
The process involves performing harmonic decomposition of perturbation equations to obtain the wave modes ($k$). Then the dependence on $k$ of energy density gradient variable
is used to obtain the transfer function which can be related to matter power spectrum. The matter power spectrum is computed at a fixed
redshift, say today.

\noindent This paper is organised as follows. In the next section, we provide and defined dimensionless variables and their evolution
equations. In Section \ref{PERTUURBATION111}, we provide perturbation equations. Section \ref{POWERSPECTRUM} is about computation
of power spectrum after making perturbation equations dimensionless. In Sections \ref{DISCUSSIONS} and \ref{CONCLUSIONS}, we provide discussions and conclusions respectively. The adopted spacetime signature is (-+++) and unless stated otherwise, we have
used the convention $8\pi G = c = 1$, where $G$ is the gravitational constant and $c$ is
the speed of light.

\section{The Theoretical Framework}\label{Framework}
\noindent The action for scalar-tensor theory (a Brans-Dicke sub-class) is given as \cite{scalar5,scalar1}
\begin{equation}
I_{f(\phi)}=\frac{1}{2}\int d^{4}x\sqrt{-g}\left[f(\phi)+\mathcal{L}_{m}\right]\; ,\label{actionST}
\end{equation}
where $R$ is the Ricci scalar and $\mathcal{L}_{m}$ is the matter Lagrangian. 
The action that represents $f(R)$ gravity given as
\begin{equation}
I=\frac{1}{2}\int d^{4}x\sqrt{-g}\left[f(R)+\mathcal{L}_{m}\right]\; .
\end{equation}
We can connect the two actions by defining the scalar field $\phi$ to be function of Ricci-scalar as \cite{ntahompagaze2017f,scalar5}
\begin{equation}\label{fphi}
\phi=f'-1\; .
\end{equation}
Here, the scalar field $\phi$ should be invertible \cite{scalar1,scalar3}. One can see that for GR case, this scalar field vanishes.
For $f(R)$ gravity, we have the curvature fluid thermodynamic quantities for an FRLW universe as background 
for zero-order are here given as
\begin{eqnarray}
&&\mu_{R}=\frac{1}{f'}\left[\frac{1}{2}(Rf'-f)-\Theta f''\dot{R}\right]\; ,\\
&&p_{R}=\frac{1}{f'}\left[\frac{1}{2}(f-Rf')+f''\ddot{R}+f'''\dot{R}^{2}+\frac{2}{3}\Theta f''\dot{R}\right]\; , 
\end{eqnarray}
where $\mu_{R}$ is the curvature energy density and $p_{R}$ is the isotropic pressure of the curvature fluid, here $\theta$ is the volume expansion parameter.
The trace equation leads to
\begin{equation}
\frac{\mu_{m}}{f'}-\frac{3p_{m}}{f'}+R-\frac{2f}{f'}-\frac{9Hf''\dot{R}}{f'}-\frac{3f''\ddot{R}}{f'}-\frac{3f'''\dot{R}^{2}}{f'} =0\; ,
\label{tracesyst1}
\end{equation}
where $\mu_{m}$ is the matter energy density and $p_{m}$ is isotropic pressure of matter fluid. The Ricci scalar is given as \cite{abebe2013large}
\begin{equation}
 R=6H^{2}\Big(\frac{\dot{H}}{H^{2}}+2\Big)\; .
\end{equation}
The Friedmann equation is given as
\begin{equation}
H^{2}=\frac{R}{6}-\frac{f}{6f'}-\frac{\dot{R}Hf''}{f'}+\frac{\mu_{m}}{3f'}-\frac{k}{a^{2}}\; , \label{dynamicall0}
\end{equation}
where we have used the fact that $H=\frac{1}{3}\Theta$.
The Raychaudhuri equation is given as
\begin{equation}
\dot{H}=-H^{2}-\frac{1}{6f'}\left(2\mu_{m}-f-6\dot{R}Hf''\right)\; . \label{dynamicall1}
\end{equation}
\subsection{The dynamical system of Radiation-dust mixture}\label{DYNAMUCAL1}
We can rewrite the Friedmann equation $(\ref{dynamicall0})$ as
\begin{equation}
1-\frac{R}{6H^{2}}+\frac{f}{6f'H^{2}}+\frac{\dot{R}f''}{f'H}-\frac{\mu_{r}}{3f'H^{2}} -\frac{\mu_{d}}{3f'H^{2}}=0\; , 
\end{equation}
where we have considered flat universe and the matter energy density $\mu_{m}$ has been decomposed into its components of
dust energy density $\mu_{d}$ and radiation energy density $\mu_{r}$. We define dimensionless parameters  \cite{26,abebe2013large}
\begin{equation}
x\equiv \frac{\dot{R}f''}{f'H}\;, \quad y\equiv\frac{f}{6f'H^{2}}-\frac{R}{6H^{2}}\;,\quad \Omega_{r}\equiv\frac{\mu_{r}}{3f'H^{2}}\;, \quad \Omega_{d}\equiv\frac{\mu_{d}}{3f'H^{2}}\;,
\end{equation}
so that the Friedmann equation can be re-expressed in dimensionless parameters as
\begin{equation}
1+y+x-\Omega_{d}-\Omega_{r}=0\; . \label{constraintone}
\end{equation}
One of the simplest and well studied models of $f(R)$ gravity theory is the power law model \cite{abebe2013large}
\begin{equation}
f(R)=\beta H^{2}_{0} \Big(\frac{R}{H^{2}_{0}}\Big)^{n}\;.
\end{equation}
 In this model, the 
the dimensionless variables can be reduced as \cite{26}:
\begin{equation}
x=\frac{\dot{R}(n-1)}{RH}\;, \quad y=\frac{R(1-n)}{6nH^{2}}\;, \quad \Omega_{r}=\frac{\mu_{r}}{3n\beta H^{2}\Big(\frac{R}{H^{2}_{0}}\Big)^{n-1}}\;, \quad \Omega_{d}=\frac{\mu_{d}}{3n\beta H^{2}\Big(\frac{R}{H^{2}_{0}}\Big)^{n-1}}\;. \label{dimeslessssssssss}
\end{equation}
From equation $(\ref{dynamicall1})$, we have
\begin{equation}
\frac{\dot{H}}{H^{2}}=-1-(\Omega_{r}+\Omega_{d})-\frac{y}{n-1}+x. \label{dotH1}
\end{equation}
Let us now define a  logarithmic time variable as 
\begin{equation}
N=\ln (a) \;,
\end{equation}
with $a$ being the scale factor, such that the evolution of the dimensionless variables $x,y,\Omega_{d}$ and $\Omega_{r}$ in Eq. \eqref{dimeslessssssssss} can be given as  \cite{26}
\begin{eqnarray}
&&\frac{dx}{dN}= -x-x^{2}+\frac{(4-2n+nx)y}{n-1}+\Omega_{d}\; ,\label{xN} \\
&&\frac{dy}{dN}=4y+\frac{(x+2ny)y}{n-1}\; ,\label{yN}\\
&&\frac{d\Omega_{d}}{dN}=\Big(1-x+\frac{2ny}{n-1}\Big)\Omega_{d}\; , \label{OmegadN}\\
&&\frac{d\Omega_{r}}{dN}= \Big(-x+\frac{2ny}{n-1}\Big)\Omega_{r}\; .\label{OmegarN}
\end{eqnarray}
The stability of the equilibrium points of the above system has been studied in details in \cite{26}. We can write these evolution equations
as functions of redshift $z$. To this end, we start from the relationship between the cosmological scale factor $a$  and 
cosmological redshift $z$ given as \cite{davies1992new,mukhanov2005physical}
$a=\frac{1}{1+z}$,
so that for any quantity $X(z)$, we can write
$\frac{dX}{dN}=-(1+z)\frac{dX}{dz}$.
We therefore have equations $(\ref{xN}), (\ref{yN})$, $(\ref{OmegarN})$ and $(\ref{OmegadN})$ given in terms of redshift $z$ as
\cite{abebe2013large}
\begin{eqnarray}
&&-(1+z)\frac{dx}{dz}=-x-x^{2}+\frac{(4-2n+nx)y}{n-1}+\Omega_{d}\; ,\label{xz} \\
&&-(1+z)\frac{dy}{dz}=4y+\frac{(x+2ny)y}{n-1}\; , \label{yz}\\
&&-(1+z)\frac{d\Omega_{d}}{dz}=\Big(1-x+\frac{2ny}{n-1}\Big)\Omega_{d} \; , \label{Omegadz}\\
&&-(1+z)\frac{d\Omega_{r}}{dz}=\Big(-x+\frac{2ny}{n-1}\Big)\Omega_{r}\; . \label{Omegarz}
\end{eqnarray}
If one defines $h=\frac{H}{H_{0}}$ \cite{abebe2013large}, then its evolution given as
\begin{equation}
(1+z)\frac{dh}{dz}=\frac{h(2+ny)}{(n-1)}\; . \label{evolutionofh} 
\end{equation}

\section{Perturbation equations}\label{PERTUURBATION111}
In the language of the scalar-tensor theory, we write linear quantities of curvature energy density and pressure as \cite{ntahompagaze2020multifluid}
\begin{eqnarray}
&& \mu_{\phi}=\frac{1}{\phi+1}\left[\frac{1}{2}\Big((\phi+1)R-f\Big)-\Theta\dot{\phi}-\phi'\tilde{\nabla}^{2}R\right]\; ,\\
&&p_{\phi}=\frac{1}{\phi+1}\left[\frac{1}{2}\Big((f-R(\phi+1)\Big)+\ddot{\phi}-\frac{\dot{\phi}\dot{\phi}'}{\phi'}\phi''
+\frac{\phi''\dot{\phi}^{2}}{\phi'^{2}}+\frac{2}{3}\Theta \dot{\phi}-\frac{2\phi'\tilde{\nabla}^{2}R}{3}\right]\; . 
\end{eqnarray}
We follow the $1+3$ covariant approach to define gradient variables to be used to build perturbation equations. This approach allows decomposition of the four-dimensional cosmological manifold into a three dimensional sub-manifold perpendicular to a timelike
vector field $u^{a}$. The derivatives (temporal and spatial) 
follow \cite{ellis1989covariant,bruni1992cosmological,Ellisbook1,amare4}.  
The scalar components of the perturbations are believed to be the ones responsible for most of the large-scale structure formation. We thus extract the scalar parts from the vector gradient quantities using the local decomposition for a vector $X_{a}$ as \cite{amare4,27}
\begin{equation}
a\tilde{\nabla}_{b}X_{a}=X_{ab}=\frac{1}{3}h_{ab}X+\Sigma^{X}_{ab}+X_{[ab]}\; , 
\end{equation}
where $\Sigma^{X}_{ab}=X_{(ab)}-\frac{1}{3}h_{ab}X$ describes shear and $X_{[ab]}$ describes vorticity. We therefore define 
the gradient variables as 
\begin{equation}
\Delta_{m}=\frac{a^{2}}{\mu_{m}}\tilde{\nabla}^{2}\mu_{m}\; , Z=a^{2}\tilde{\nabla}^{2}\Theta \; , 
\Phi =a^{2}\tilde{\nabla}^{2}\phi\; , 
\text{ and }
\Psi =a^{2}\tilde{\nabla}^{2}\dot{\phi} \; . 
\end{equation}
We use harmonic decomposition method to transform covariant Laplace-Beltrami operator \cite{bruni1992cosmological,Ellisbook1,amare4}.
For a given quantity $X$, we write
\begin{equation}
X=\sum_{k}X^{k}(t)Q_{k}(\vec{x}), 
\end{equation}
where $Q_{k}$ are the eigenfunctions of the covariant Laplace-Beltrami operator such that
\begin{equation}
 \tilde{\nabla}^{2}Q=-\frac{k^{2}}{a^{2}}Q\; ,
\end{equation}
and the order of the scale factor dependent harmonic (wavenumber) $k$ is 
\begin{equation}
 k=\frac{2\pi a}{\lambda}\; , \label{wavenumberk}
\end{equation}
where $\lambda$ is the physical wavelength of the mode and a is the scale factor. 
The evolution of the above gradient variables leads to perturbation equations given as
\begin{eqnarray}
 \ddot{\Delta}^{k}_{m}&-&\Big[\Big(\frac{\dot{\phi}}{\phi+1}-\frac{2\Theta}{3}\Big)+w\Theta\Big]\dot{\Delta}^{k}_{m}
-\Big[-w\Theta\Big(\frac{\dot{\phi}}{\phi+1}-\frac{2\Theta}{3}\Big)+\frac{w+1}{(\phi+1)}\mu_{m}\nonumber\\
&-&w\Big(\mu_{m}+\frac{1}{3}\Theta^{2}-\frac{f}{2(\phi+1)}-\frac{k^{2}\phi'R}{a^{2}(\phi+1)}\Big)
+w\dot{\Theta} -\frac{wk^{2}}{a^{2}}\Big]\Delta^{k}_{m}+\frac{(1+w)\Theta}{\phi+1}\dot{\Phi}^{k}\nonumber\\
&+&(1+w)\Big[\frac{1}{2\phi'}+\frac{2\mu_{m}-f-2\Theta\dot{\phi}}{2(\phi+1)^{2}}+\frac{k^{2}\phi''R}{a^{2}\phi'(\phi+1)}
 -\frac{k^{2}\phi'R}{a^{2}(\phi+1)^{2}} +\frac{k^{2}}{a^{2}(\phi+1)} \Big]\Phi^{k}=0 \; ,\;\;\;\;\;\;\;\\
\ddot{\Phi}^{k}&-&\Big[\frac{\ddot{\phi}\phi''}{\phi'^{2}}-\frac{2\dot{\phi}\dot{\phi}'\phi''}{\phi'^{3}}-\frac{(\phi+1)}{3\phi'}+\frac{R}{3} -\frac{R(\phi+1)\phi''}{3\phi'^{2}} 
-\frac{(1-3w)\phi''\mu_{m}}{3\phi'^{2}}
+\frac{2f\phi''}{3\phi'^{2}} -\frac{\dot{R}^{2}\phi'''}{\phi'} \;\;\nonumber\\
&+&\frac{\Theta\phi''\dot{\phi}}{\phi'^{2}}
+\frac{3\dot{\phi}^{2}\phi''^{2}}{\phi'^{4}}
+\frac{\dot{\phi}\dot{\phi}''}{\phi'^{2}}-\frac{k^{2}}{a^{2}}\Big]\Phi^{k} 
-\Big(\frac{\dot{\phi}'}{\phi'}-\Theta-\frac{2\dot{\phi}\phi''}{\phi'^{2}}\Big)\dot{\Phi}^{k}
-\Big[\frac{w\dot{\phi}}{w+1}\Big(\frac{\dot{\phi}'}{\phi'}-\frac{2\dot{\phi}\phi''}{\phi'^{2}}\Big)\nonumber\\
&-&\frac{2w\ddot{\phi}}{(w+1)} +\frac{(1-3w)\mu_{m}}{3} -\frac{2w\Theta\dot{\phi}}{1+w}\Big]\Delta^{k}_{m} 
-\Big(\frac{\dot{\phi}}{1+w}-\frac{w\dot{\phi}}{(w+1)}\Big)\dot{\Delta}^{k}_{m}=0 \;.
\end{eqnarray}

\noindent Note that we have instituted the harmonic decomposition along the way. As most of the large-scale structures are believe to have formed  during the epoch of matter domination, this epoch rules mainly after decoupling. We therefore consider the dust case with $w=0$ so that we write the above equations as
\begin{eqnarray}
\ddot{\Delta}^{k}_{d}&-&\Big(\frac{\dot{\phi}}{\phi+1}-\frac{2\Theta}{3}\Big)\dot{\Delta}^{k}_{d}
-\frac{\mu_{d}}{(\phi+1)}\Delta^{k}_{d}+\frac{\Theta}{\phi+1}\dot{\Phi}^{k}
+\Big[\frac{1}{2\phi'}+\frac{2\mu_{d}-f-2\Theta\dot{\phi}}{2(\phi+1)^{2}}\nonumber\\
&+&\frac{k^{2}\phi''R}{a^{2}\phi'(\phi+1)}
-\frac{k^{2}\phi'R}{a^{2}(\phi+1)^{2}} +\frac{k^{2}}{a^{2}(\phi+1)} \Big]\Phi^{k}=0 \; ,\label{beforedynamicaltransformationDelta}\\
\ddot{\Phi}^{k}&-&\Big[\frac{\ddot{\phi}\phi''}{\phi'^{2}}-\frac{2\dot{\phi}\dot{\phi}'\phi''}{\phi'^{3}}
-\frac{(\phi+1)}{3\phi'}+\frac{R}{3} -\frac{R(\phi+1)\phi''}{3\phi'^{2}} 
-\frac{\phi''\mu_{d}}{3\phi'^{2}}
+\frac{2f\phi''}{3\phi'^{2}} -\frac{\dot{R}^{2}\phi'''}{\phi'} \nonumber\\
&+&\frac{\Theta\phi''\dot{\phi}}{\phi'^{2}}
+\frac{3\dot{\phi}^{2}\phi''^{2}}{\phi'^{4}}
+\frac{\dot{\phi}\dot{\phi}''}{\phi'^{2}}-\frac{k^{2}}{a^{2}}\Big]\Phi^{k} 
-\Big(\frac{\dot{\phi}'}{\phi'}-\Theta-\frac{2\dot{\phi}\phi''}{\phi'^{2}}\Big)\dot{\Phi}^{k}
-\frac{\mu_{d}}{3}\Delta^{k}_{d} 
-\dot{\phi}\dot{\Delta}^{k}_{d}=0 \;\;\;\;\;\;\;\;\;.\label{beforedynamicaltransformationPhi}
\end{eqnarray}
To easily analyse the above equations, we transform them into the dimensionless form. Each term was transformed such a way that we have the perturbations equations written as
\begin{eqnarray}
\Delta^{''k}_{d}&+&\Big[\frac{(2+ny)}{(1+z)(n-1)}
+\frac{x-1}{(1+z)}\Big]\Delta^{'k}_{d}
-\frac{3\Omega_{d}}{(1+z)^{2}}\Delta^{k}_{d}-\frac{3(1-n)^{n-1}}{(1+z)\beta n (6nh^{2}y)^{n-1}}\Phi^{'k}\nonumber\\
&+&\frac{1}{(1+z)^{2}}\Big[\frac{(1-n)^{n-2}}{2h^{2}\beta n(n-1)(6nh^{2}y)^{n-2}}
+\frac{3(1-n)^{n-1}}{\beta n (6nh^{2}y)^{n-1}}\Big(\Omega_{d}-x-\frac{y}{n-1}\Big)\Big]\Phi^{k}=0\;,\;\;\;\;\;\;\;\label{Deltaafterdyanmical1}\\
\Phi^{''k}&+&\Big[\frac{(2+ny)}{(1+z)(n-1)}
-\frac{2}{(1+z)} -\frac{x(n-2)}{(n-1)((1+z))}\Big]\Phi^{'k} 
-\frac{(n-2)}{(1+z)^{2}}\Big[-\frac{\Omega_{d}}{n-1}\nonumber\\
&+&\frac{4(1-q)}{3n(n-1)} 
+\frac{(n^{2}-3n-4)x^{2}}{(n-1)^{2}}
+\frac{3x}{(n-1)}-(n-2)x^{2}
-\frac{\hat{k}^{2}}{(n-2)}\Big]\Phi^{k} \nonumber\\
&-&\frac{\Omega_{d}\beta n (6nh^{2}y)^{n-1}}{(1+z)^{2}(1-n)^{n-1}}\Delta^{k}_{d} 
-\frac{\beta xn(6nh^{2}y)^{n-1}}{(1+z)(1-n)^{n-1}} \Delta^{'k}_{d}=0 \;,\label{Phiafterdyanmical1}
\end{eqnarray}
where 
$q=\frac{ny}{n-1}+1$.
In the following section, we use these two equations to compute the power spectrum of dust energy density perturbations.

\section{Power spectrum}\label{POWERSPECTRUM}
The matter power spectrum in modified gravity theories was widely explored in cosmology \cite{ abebe2013large, jennings2012redshift, tsujikawa2009dispersion, zhao2011n, li2011chameleon, brax2012systematic, brax2013signatures, he2013revisiting}. In the present paper, the analysis will be done in line with the work covered in  \cite{abebe2013large}. Having said that, we can define
a transfer function $T(k)$ given as \cite{abebe2013large,27}
\begin{equation}
T(k)=\left\langle \left|\Delta^{k}_{m}\right|^{2} \right\rangle\; , 
\end{equation}
where $\Delta^{k}_{m}$ is the matter density perturbation and $k$ is the wave-number that characterizes the harmonic mode. 
Due to isotropy of the observable universe, one has
\begin{equation}
 T(k)= \left|\Delta^{k}_{m}\right|^{2}\; . 
\end{equation}
The power spectrum is a quantity $P(k)$ is proportional to the transfer function $T(k)$
\begin{equation}
P(k)\varpropto T(k)\; .
\end{equation}
This is shown by the relation  \cite{abebe2013large}
\begin{equation}
P^{f(R)}_{k}=T(k)P^{\Lambda CDM}_{k}\Big|_{eq}\; , 
\end{equation}
where $P^{\Lambda CDM}_{k}\Big|_{eq}$ is the power spectrum during epoch of the equality of radiation and matter in the linear regime in the $\Lambda CDM$ model. Thus one can write power spectrum $P^{\phi}(k)$ given as
\begin{equation}
P^{\phi}(k)=\frac{P^{f(\phi)}_{k}}{P^{\Lambda CDM}_{k}\Big|_{eq}}=\left|\Delta^{k}_{d}\right|^{2}, \label{powerspeceq}
\end{equation}
where $\Delta^{k}_{d}$ is the pressureless dust (baryonic matter) energy density perturbation obtained after solving the system of equations \eqref{Deltaafterdyanmical1}
and \eqref{Phiafterdyanmical1}. To obtain the results, we have simultaneously solved Eqs. \eqref{xz}-\eqref{evolutionofh} together with the constraint Eq. \eqref{constraintone} to obtain the redshift dependent solutions for parameters $x(z),y(z)$ as well as $\Omega_{d}(z)$. Having obtained such parameters, with the consideration of different values of  $n$, one solved simultaneously two equations namely Eq. \eqref{Deltaafterdyanmical1} and Eq. \eqref{Phiafterdyanmical1} to obtain $\Delta^{k}_{d}(z)$ to be used in the plotting of the Eq. \eqref{powerspeceq} with different set of initial conditions. The plotting was done at a fixed redshift $(z=0)$, means today and the power spectrum $P(k)$ was plotted against $k$. To easily visualize the changes and behaviors, the plots are presented with log on both axes.  

\begin{figure}[H]
\centering
\includegraphics[scale=0.5]{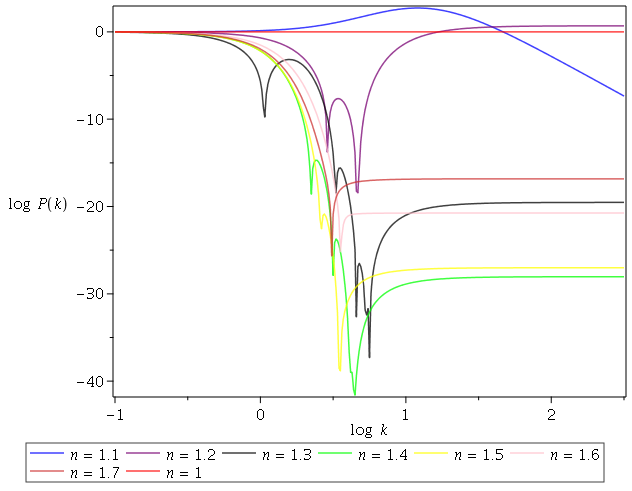}
\caption{Power spectrum for $n=1,n=1.1,, n=1.2, n=1.3, n=1.4, n=1.5, n=1.6, n=1.7$, initial conditions at $z_{0}=2000$ 
are $\Delta(z_{0})=10^{-5}$, $\Delta'(z_{0})=10^{-5}$,
$\Phi(z_{0})=10^{-5}$ and $\Phi'(z_{0})=10^{-5}$, obtained from Eq. \eqref{powerspeceq}.}
\label{Figspectrumdifferent0}
\centering
\includegraphics[scale=0.45]{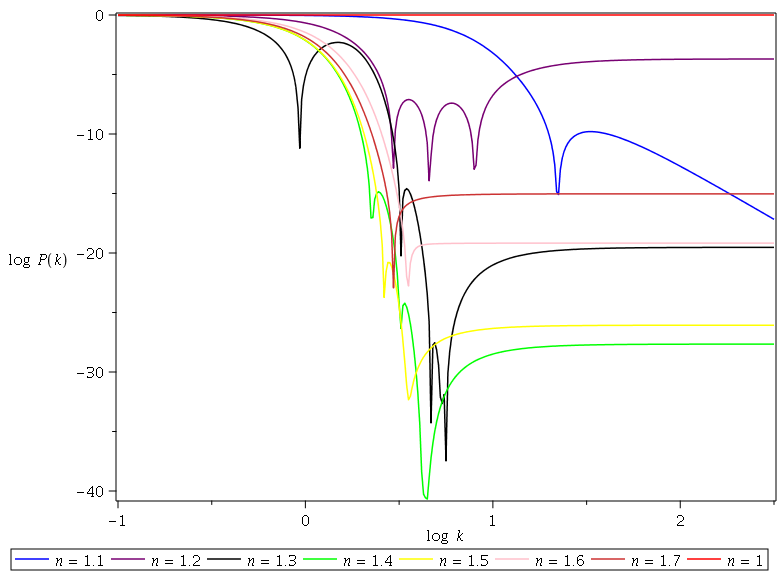}
\caption{Power spectrum for $n=1,n=1.1,, n=1.2, n=1.3, n=1.4, n=1.5, n=1.6, n=1.7$, initial conditions at $z_{0}=2000$ 
are $\Delta(z_{0})=10^{-5}$, $\Delta'(z_{0})=0$,
$\Phi(z_{0})=10^{-5}$ and $\Phi'(z_{0})=0$, obtained from Eq. \eqref{powerspeceq}.}
\label{Figspectrumdifferent1}
\end{figure}
\section{Discussions}\label{DISCUSSIONS}
\begin{figure}[H]
\centering
\includegraphics[scale=0.45]{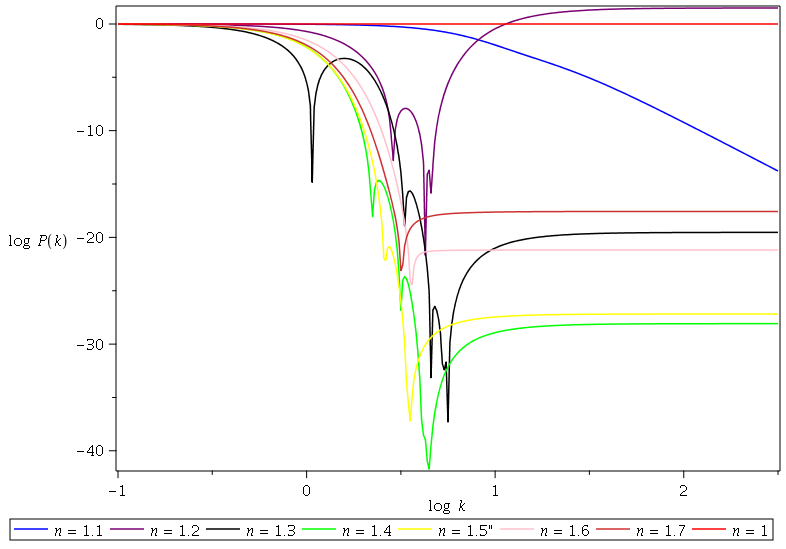}
\caption{Power spectrum for $n=1,n=1.1,, n=1.2, n=1.3, n=1.4, n=1.5, n=1.6, n=1.7$, initial conditions at $z_{0}=2000$ 
are $\Delta(z_{0})=10^{-5}$, $\Delta'(z_{0})=10^{-8}$,
$\Phi(z_{0})=10^{-5}$ and $\Phi'(z_{0})=10^{-8}$, obtained from Eq. \eqref{powerspeceq}.}
\label{Figspectrumdifferent2}
\centering
\includegraphics[scale=0.42]{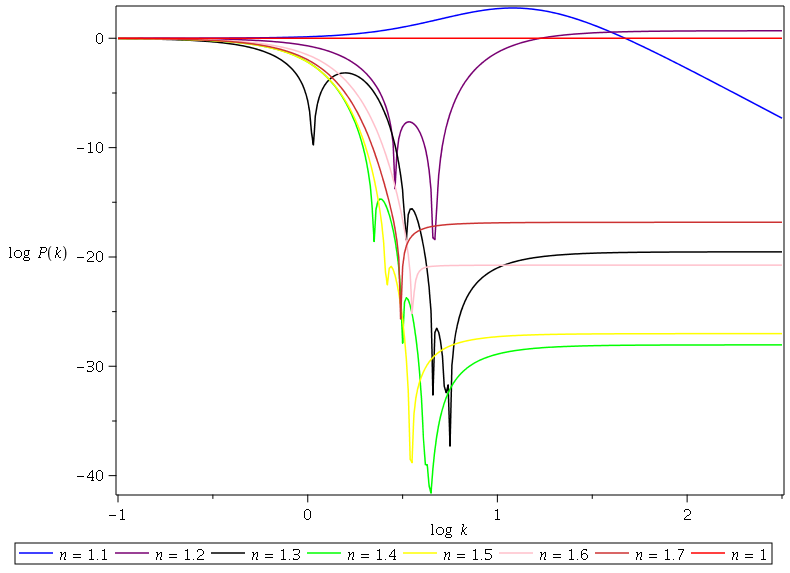}
\caption{Power spectrum for $n=1,n=1.1,, n=1.2, n=1.3, n=1.4, n=1.5, n=1.6, n=1.7$, initial conditions at $z_{0}=2000$ 
are $\Delta(z_{0})=10^{-5}$, $\Delta'(z_{0})=10^{-3}$,
$\Phi(z_{0})=10^{-5}$ and $\Phi'(z_{0})=10^{-3}$, obtained from Eq. \eqref{powerspeceq}.}
\label{Figspectrumdifferent3}
\end{figure}

\begin{figure}[H]
\centering
\includegraphics[scale=0.35]{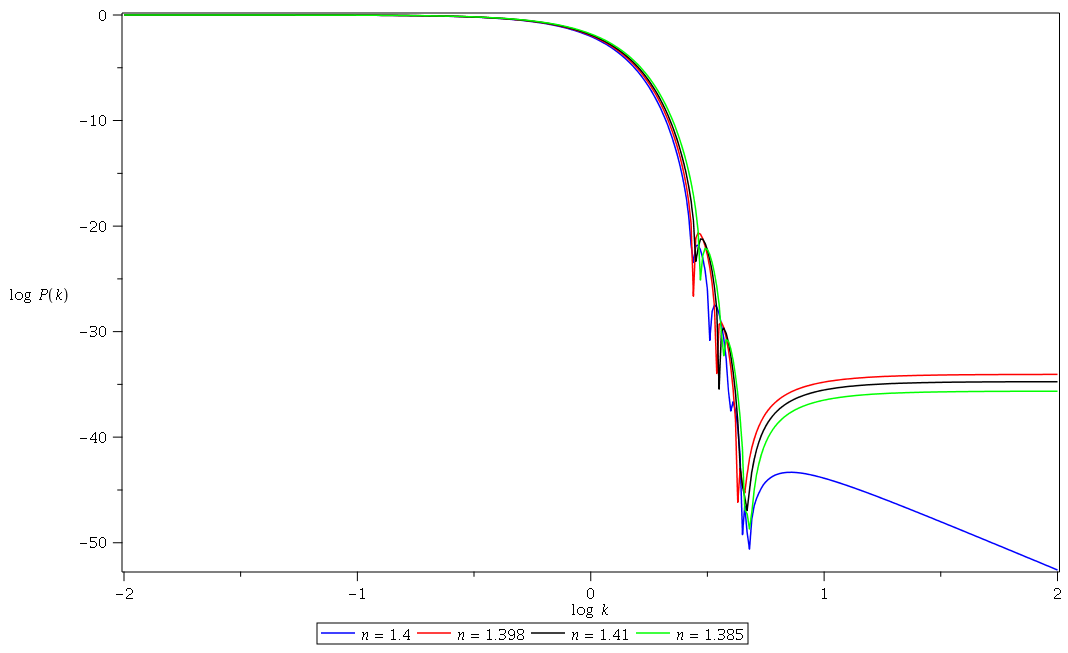}
\caption{Power spectrum for $n=1.41, n=1.4, n=1.385,n=1.398$, initial conditions at $z_{0}=2000$ 
are $\Delta(z_{0})=10^{-5}$, $\Delta'(z_{0})=10^{-5}$,
$\Phi(z_{0})=10^{-5}$ and $\Phi'(z_{0})=10^{-5}$, obtained from Eq. \eqref{powerspeceq}.}
\label{Figspectrum1}
\centering
\includegraphics[scale=0.38]{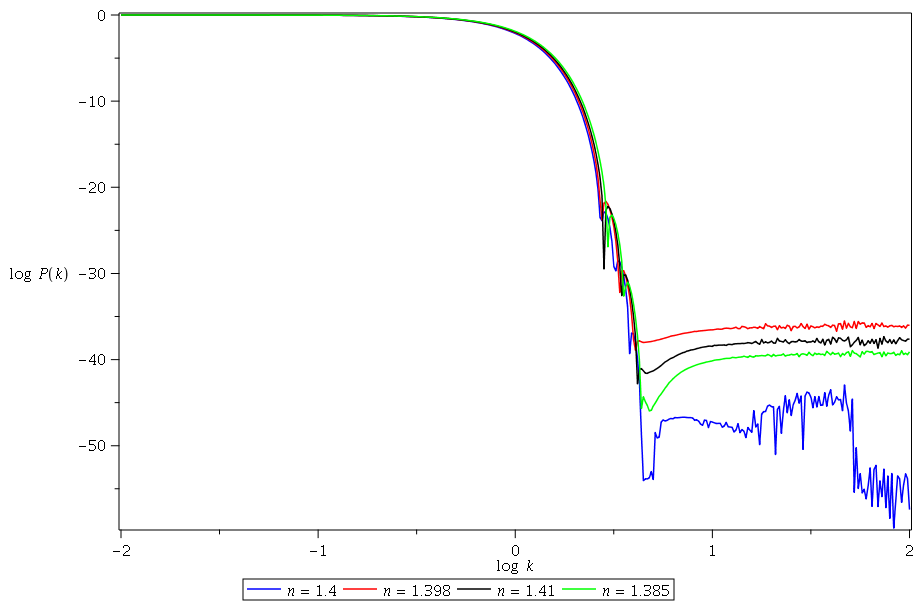}
\caption{Power spectrum for $n=1.41, n=1.4, n=1.385,n=1.398$, initial conditions at $z_{0}=2000$ 
are $\Delta(z_{0})=10^{-5}$, $\Delta'(z_{0})=0$,
$\Phi(z_{0})=10^{-5}$ and $\Phi'(z_{0})=0$, obtained from Eq. \eqref{powerspeceq}.}
\label{Figspectrum2}
\end{figure}

\begin{figure}[H]
\centering
\includegraphics[scale=0.4]{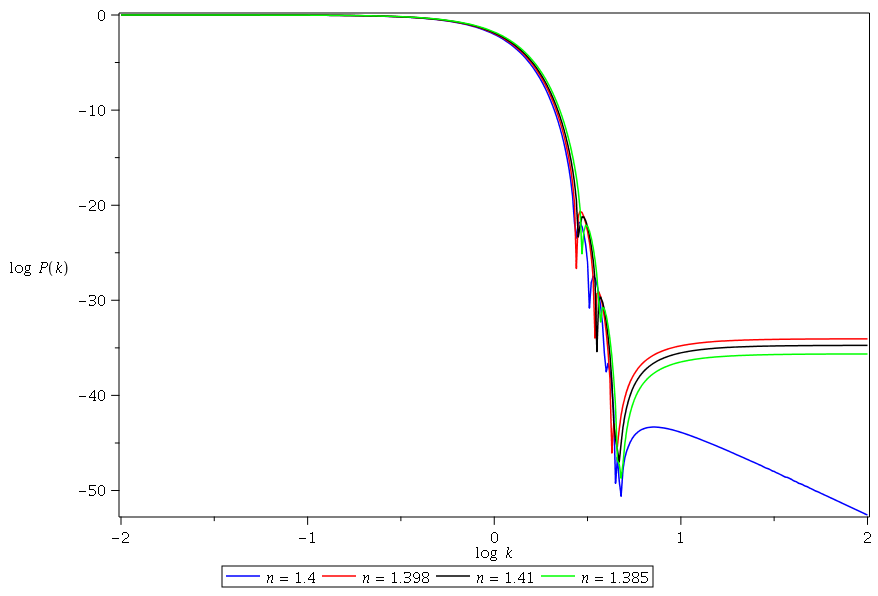}
\caption{Power spectrum for $n=1.41, n=1.4, n=1.385,n=1.398$, initial conditions at $z_{0}=2000$ 
are $\Delta(z_{0})=10^{-5}$, $\Delta'(z_{0})=10^{-8}$,
$\Phi(z_{0})=10^{-5}$ and $\Phi'(z_{0})=10^{-8}$, obtained from Eq. \eqref{powerspeceq}.}
\label{Figspectrum3}
\centering
\includegraphics[scale=0.4]{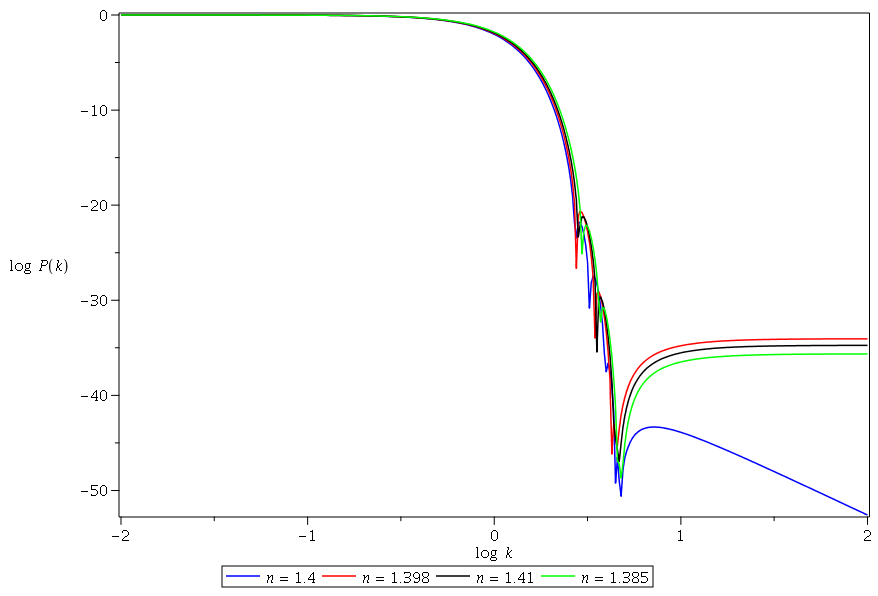}
\caption{Power spectrum for $n=1.41, n=1.4, n=1.385,n=1.398$, initial conditions at $z_{0}=2000$ 
are $\Delta(z_{0})=10^{-5}$, $\Delta'(z_{0})=10^{-3}$,
$\Phi(z_{0})=10^{-5}$ and $\Phi'(z_{0})=10^{-3}$, obtained from Eq. \eqref{powerspeceq}.}
\label{Figspectrum4}
\end{figure}

We consider different values of the power law index $n$ in $R^{n}$ model following the choice made in \cite{27} and maintain the initial conditions
used in \cite{abebe2013large}. 
The plots for matter power spectrum are presented in Fig. \ref{Figspectrumdifferent0}, Fig. \ref{Figspectrumdifferent1}, Fig. \ref{Figspectrumdifferent2} and
Fig. \ref{Figspectrumdifferent3}. 
The initial conditions are of four types. All of them are considered at redshift $z_{0}=2000$.
The first set of initial condition is $I: \Delta(z_{0})=10^{-5},\Delta'(z_{0})=10^{-5},
\Phi(z_{0})=10^{-5} \text{ and } \Phi'(z_{0})=10^{-5}$.
The second set of initial condition $II: \Delta(z_{0})=10^{-5},\Delta'(z_{0})=0,
\Phi(z_{0})=10^{-5} \text{ and } \Phi'(z_{0})=0$. The third set of initial condition is
$III: \Delta(z_{0})=10^{-5},\Delta'(z_{0})=10^{-8},
\Phi(z_{0})=10^{-5} \text{ and } \Phi'(z_{0})=10^{-8}$. The last set of initial condition is 
$IV: \Delta(z_{0})=10^{-5},\Delta'(z_{0})=10^{-3},
\Phi(z_{0})=10^{-5} \text{ and } \Phi'(z_{0})=10^{-3}$.
The effects of setting some of the above initial conditions are discussed in details in \cite{abebe2013large}.
The background equations take initial values from the observations. We have used observable parameters such as the Hubble constant $H_{0}$, deceleration
parameter $q_{0}$ and density parameters taken from $\Lambda$CDM initial conditions.
The perturbations are evaluated at redshift $z=0$ that is today.
The ranges of $k$ used are taken from \cite{tegmark2006cosmological}. The $\Lambda$CDM model's cosmological parameters are 
taken from the SDSS and Planck data Survey \cite{bahcall2003richness,collaboration2014planck} respectively.
Each type produces changes in amplitudes. The GR case is 
for $n=1$ where the power spectrum is scale invariant under dust perturbations. For $1<n<1.3$, with the first set of initial conditions
we have power spectrum evolving above GR invariant line. For $n\geq 1.3$, the power spectrum starts with constant amplitude 
then it experiences oscillations and eventually saturates at finite amplitude. 
This behavior of the power spectrum is similar to the one presented in \cite{27}  for $f(R)$ gravity, where they have obtained three different
regimes depending on the finite amplitude-saturation of the power spectrum.  
We have paid attention to $n=1.4$ since it is very clear from the power spectrum that the values close to $n=1.4$ are behaving similarly and have 
the finite amplitude.
We choose values close to $n=1.4$ to see how their corresponding power spectrum behave. We have plotted them in Fig. \ref{Figspectrum1},
Fig. \ref{Figspectrum2}, Fig. \ref{Figspectrum3} and \ref{Figspectrum4}. For the initial conditions of type $II$, the oscillations are 
frequent compared to other initial conditions during the amplitude finite-saturation. However, one can see
 that the power spectrum of $n=1.4$ is lower than others for large values of $k$ in all sets of initial conditions. The present work reveals a couple of information in different aspects $(i)$ the obtained perturbations are new in comparision with the work done in  \cite{abebe2013large}, $(ii)$ The obtained power spectra around $n=1.4$ have shown to saturate at finite amplitude as well, a feature not deeply explored in  \cite{abebe2013large}, $(iii)$ with some set of initial conditions, the behavior of power spectrum around $n=1.4$ shows oscillations during the finite-saturation state. This characteristics was not identified in the work done in  \cite{abebe2013large}.


\section{Conclusions}\label{CONCLUSIONS}
We have used the $1+3$ covariant approach to obtain the evolution equations and studied
the behavior of matter power spectrum of perturbation equations. These equations are written in dimensionless form using a
dynamical system approach. We have defined the gradient variables based on the equivalence between $f(R)$ theory of gravity and
scalar-tensor theory of gravity. The background cosmological parameters are considered from the known values in the $\Lambda$CDM model. We show that the behavior of the 
power spectrum obtained is similar to the one obtained in \cite{27}, the authors focused on $f(R)$ theory only.
We have extended the discussion to scalar tensor theory using a dynamical system to treat linear perturbations and compute matter power spectrum. In the present work, we obtained newly equations based on the gradients developed that involve the definition of the scalar field provided.\\
\noindent We noted that for $1<n<1.3$, with the first set of initial conditions
we have power spectrum evolving above the GR scale-invariant line. For $n\geq 1.3$, the power spectrum starts with constant amplitude 
then evolves into oscillations and eventually saturates at a finite amplitude. 
This behavior of the power spectrum is similar to the one presented in \cite{27}  for $f(R)$ gravity, where authors obtained three different
regimes depending on the finite amplitude-saturation of the power spectrum.  
We have paid attention to $n=1.4$ since it is very clear from the power spectrum that the values close to $n=1.4$ are behaving similarly and have 
the finite amplitude.
We choose values close to $n=1.4$ to study how their corresponding power spectra behave. We have plotted the results in Fig. \ref{Figspectrum1},
Fig. \ref{Figspectrum2}, Fig. \ref{Figspectrum3} and \ref{Figspectrum4}. For the initial conditions of type $II$, the oscillations are 
frequent compared to other initial conditions during the amplitude finite-saturation, a feature not previously explored in  \cite{abebe2013large}. However, one can see
 that the power spectrum of the  $n=1.4$ case is lower than the others for large values of $k$ in all sets of initial conditions. Such behaviors were  previously observed in the literature as well \cite{27}. The present results, where there is a consideration of the definition of scalar field that vanishes in GR limit, support the ongoing investigations of the equivalence between $f(R)$ theory and scalar-tensor theory at linear order.
\section*{Acknowledgments}  
AA acknowledge that this work is based on the research supported in part by the National Research Foundation (NRF) of South Africa (grant number 112131).  AA also acknowledges the hospitality of the Department of Physics of the University of Rwanda, where  part of this work was completed. JN and MM acknowledge the support from International Science Program (ISP) to Rwanda Astrophysics, Space and Climate Science Research Group (RASCSRG), University of Rwanda (grant number RWA01).


\begin{thebibliography}{10}

\bibitem{dunsby1992covariant}
Peter~KS Dunsby, Marco Bruni, and George~FR Ellis.
\newblock Covariant perturbations in a multifluid cosmological medium.
\newblock {\em The Astrophysical Journal}, 395:54--74, 1992.

\bibitem{arbuzov2009general}
AB~Arbuzov et~al.
\newblock General relativity and the standard model in scale-invariant
  variables.
\newblock {\em Gravitation and Cosmology}, 15(3):199--212, 2009.

\bibitem{geng2015matter}
Chao-Qiang Geng, Chung-Chi Lee, and Jia-Liang Shen.
\newblock Matter power spectra in viable {$f(R)$} gravity models with massive
  neutrinos.
\newblock {\em Physics Letters B}, 740:285--290, 2015.

\bibitem{li2012non}
Baojiu Li et~al.
\newblock The non-linear matter and velocity power spectra in {$f(R)$} gravity.
\newblock {\em Monthly Notices of the Royal Astronomical Society},
  428(1):743--755, 2012.

\bibitem{li2011chameleon}
Yin Li and Wayne Hu.
\newblock Chameleon halo modeling in {$f(R)$} gravity.
\newblock {\em Physical Review D}, 84(8):084033, 2011.

\bibitem{SanteCarloni1}
Sante Carloni.
\newblock Covariant gauge invariant theory of scalar perturbations in
  ${f(R)}$-gravity: A brief review.
\newblock {\em Open Astronomy Journal}, 3:76--93, 2010.

\bibitem{abebe2015breaking}
Amare Abebe.
\newblock Breaking the cosmological background degeneracy by two-fluid
  perturbations in {$f(R)$} gravity.
\newblock {\em International Journal of Modern Physics D}, 24(07):1550053,
  2015.

\bibitem{ntahompagaze2017f}
Joseph Ntahompagaze, Amare Abebe, and Manasse Mbonye.
\newblock On ${f(R)}$ gravity in scalar--tensor theories.
\newblock {\em International Journal of Geometric Methods in Modern Physics},
  14(07):1750107, 2017.

\bibitem{26}
Sante Carloni, Peter~KS Dunsby, Salvatore Capozziello, and Antonio Troisi.
\newblock Cosmological dynamics of ${R}^{n}$ gravity.
\newblock {\em Classical and Quantum Gravity}, 22(22):4839, 2005.

\bibitem{bahamonde2018dynamical}
Sebastian Bahamonde, Christian~G B{\"o}hmer, Sante Carloni, Edmund~J Copeland,
  Wei Fang, and Nicola Tamanini.
\newblock Dynamical systems applied to cosmology: dark energy and modified
  gravity.
\newblock {\em Physics Reports}, 775:1--122, 2018.

\bibitem{abebe2013large}
Amare Abebe, {\'A}lvaro de~la Cruz-Dombriz, and Peter~KS Dunsby.
\newblock Large scale structure constraints for a class of ${f(R)}$ theories of
  gravity.
\newblock {\em Physical review D}, 88(4):044050, 2013.

\bibitem{scalar5}
Andrei~V Frolov.
\newblock Singularity problem with ${f(R)}$ models for dark energy.
\newblock {\em Physical Review letters}, 101(6):061103, 2008.

\bibitem{scalar1}
Timothy Clifton et~al.
\newblock Modified gravity and cosmology.
\newblock {\em Physics Reports}, 513(1):1--189, 2012.

\bibitem{scalar3}
Thomas Faulkner, Max Tegmark, Emory~F Bunn, and Yi~Mao.
\newblock Constraining ${f(R)}$ gravity as a scalar-tensor theory.
\newblock {\em Physical Review D}, 76(6):063505, 2007.

\bibitem{davies1992new}
Paul Davies.
\newblock {\em The new physics}.
\newblock Cambridge University Press, 1992.

\bibitem{mukhanov2005physical}
Viatcheslav Mukhanov.
\newblock {\em Physical foundations of cosmology}.
\newblock Cambridge University Press, 2005.

\bibitem{ntahompagaze2020multifluid}
Joseph Ntahompagaze, Shambel Sahlu, Amare Abebe, and Manasse~R Mbonye.
\newblock On multifluid perturbations in scalar--tensor cosmology.
\newblock {\em International Journal of Modern Physics D}, 29(16):2050120,
  2020.

\bibitem{ellis1989covariant}
George~FR Ellis and Marco Bruni.
\newblock Covariant and gauge-invariant approach to cosmological density
  fluctuations.
\newblock {\em Physical Review D}, 40(6):1804, 1989.

\bibitem{bruni1992cosmological}
Marco Bruni, Peter~KS Dunsby, and George~FR Ellis.
\newblock Cosmological perturbations and the physical meaning of
  gauge-invariant variables.
\newblock {\em The Astrophysical Journal}, 395:34--53, 1992.

\bibitem{Ellisbook1}
George~FR Ellis, Roy Maartens, and Malcolm~AH MacCallum.
\newblock {\em Relativistic cosmology}.
\newblock Cambridge University Press, 2012.

\bibitem{amare4}
Amare Abebe et~al.
\newblock Covariant gauge-invariant perturbations in multifluid ${f(R)}$
  gravity.
\newblock {\em Classical and Quantum Gravity}, 29(13):135011, 2012.

\bibitem{27}
Kishore~N Ananda, Sante Carloni, and Peter~KS Dunsby.
\newblock Structure growth in ${f(R)}$ theories of gravity with a dust equation
  of state.
\newblock {\em Classical and Quantum Gravity}, 26(23):235018, 2009.

\bibitem{tegmark2006cosmological}
Max Tegmark et~al.
\newblock Cosmological constraints from the sdss luminous red galaxies.
\newblock {\em Physical Review D}, 74(12):123507, 2006.

\bibitem{bahcall2003richness}
Neta~A Bahcall et~al.
\newblock The richness-dependent cluster correlation function: Early {Sloan
  Digital Sky Survey} data.
\newblock {\em The Astrophysical Journal}, 599(2):814, 2003.

\bibitem{collaboration2014planck}
Planck Collaboration et~al.
\newblock Planck 2013 results. xvi. cosmological parameters.
\newblock {\em Astronomy and Astrophysics}, 571:A16, 2014.

\bibitem{jennings2012redshift}
Jennings, Elise and Baugh, Carlton M and Li, Baojiu and Zhao, Gong-Bo and Koyama, Kazuya
 \newblock Redshift-space distortions in ${f(R)}$  gravity.
 \newblock {\em Monthly Notices of the Royal Astronomical Society}, 425(3): 2128--2143, 2012.

\bibitem{tsujikawa2009dispersion}
Tsujikawa, Shinji and Gannouji, Radouane and Moraes, Bruno and Polarski, David
\newblock Dispersion of growth of matter perturbations in ${f(R)}$  gravity.
\newblock {\em Physical Review D}, 80(8):084044, 2009.
 
\bibitem{zhao2011n}
Zhao, Gong-Bo and Li, Baojiu and Koyama, Kazuya
\newblock N-body simulations for ${f(R)}$  gravity using a self-adaptive particle-mesh code.
\newblock {\em Physical Review D}, 83(4):044007, 2011.
 

\bibitem{brax2012systematic}
Brax, Philippe and Davis, Anne-Christine and Li, Baojiu and Winther, Hans A and Zhao, Gong-Bo
\newblock Systematic simulations of modified gravity: symmetron and dilaton models.
\newblock {\em Journal of Cosmology and Astroparticle Physics},2012(10):002, 2012.

\bibitem{brax2013signatures}
Brax, Philippe and Clesse, S{\'e}bastien and Davis, Anne-Christine
 \newblock Signatures of modified gravity on the 21 cm power spectrum at reionisation.
 \newblock {\em Journal of Cosmology and Astroparticle Physics}, 2013(01):003, 2013.

\bibitem{he2013revisiting}
He, Jian-hua and Li, Baojiu and Jing, YP
\newblock Revisiting the matter power spectra in ${f(R)}$  gravity.
  \newblock  {\em Physical Review D}, 88(10):103507, 2013.

\bibitem{taruya2014regularized}
Taruya, Atsushi and Nishimichi, Takahiro and Bernardeau, Francis and Hiramatsu, Takashi and Koyama, Kazuya
 \newblock Regularized cosmological power spectrum and correlation function in modified gravity models.
  \newblock {\em Physical Review D}, 90(12):123515, 2014.

\bibitem{koivisto2006matter}
Koivisto, Tomi
\newblock Matter power spectrum in ${f(R)}$  gravity.
 \newblock {\em Physical Review D}, 73(8):083517, 2006.
 



\end{thebibliography}
\bibliographystyle{unsrt}

\end{document}